# Multiferroic Studies on $La_{0.7}Bi_{0.3}CrO_3$ Perovskite

Aga Shahee, Dhirendra kumar, N. P. Lalla[#]

*UGC-DAE Consortium for Scientific Research, University campus, Khandwa Road, Indore-452001*
*[#]nplalla@csr.res.in*

**Abstract.** Magnetic and dielectric properties on a polycrystalline $La_{0.7}Bi_{0.3}CrO_3$ have been investigated. A canted antiferromagnetic (CAF) phase transition giving weak ferromagnetism at $T_N = 230K$ has been observed. M-H at 10K shows non-saturating trend, even up to 5T, confirms the CAF ordering. The permittivity (ε′)-vs-temperature (T) variation shows a relaxor ferroelectric (RFE) nature accompanied by a pronounced anomaly in ε′-T at $T_N$. The anomaly in ε′-T at $T_N$ indicates the presence of magentoelectric (ME) coupling in this material.

## INTRODUCTION

Multiferroics are materials exhibiting two or more of the ferroic orders [1]. Currently the materials having simultaneous ferromagnetic and ferroelectric orders have become of much interest. The coupling between the magnetic order and ferroelectric order is mediate by ME effect [1]. The ME effect provides an additional degree of freedom in the design of actuators and next generation memory devices. Like $BiFeO_3$, $BiCrO_3$ is also an important perovskite expected to show enhance multiferroic properties [1]. But synthesis of $BiCrO_3$ needs high temperature and high pressure. Therefore we have studied $Bi^{+3}$ doped $LaCrO_3$ which is likely to show multiferroic behaviour. In the following we have investigated the multiferroic behaviour of $La_{0.7}Bi_{0.3}CrO_3$ by probing its electrical and magnetic orders to establish ME coupling.

## EXPERIMENTAL DETIALS

Perovskite oxide $La_{0.7}Bi_{0.3}CrO_3$ was prepared by conventional solid-state reaction route using stoichiometric amounts of the starting oxides, $La_2O_3$ (99.99%), $Cr_2O_3$ (99.99%) and $Bi_2O_3$ (99.99%) (in 5% excess). The bulk structural characterization was carried out using powder XRD, which reviled the presence of single phase orthorhombic perovskite with Pnma structure. The magnetic and dielectric studies have been carried out at low temperatures (LT) using MPMS-7 SQUID magnetometer and Hioki-LCR impedance analyzer respectively. The dielectric measurement in the temperature range of 80K-390K was done using copper electrodes.

## RESULTS AND DISSCUSION

Fig.1 (a) shows 100Oe field-cool (FC) magnetization vs temperature (M-T) data from 10K-300K. It can be seen that with decreasing temperature the magnetization increases at around 230K and then it saturates. Fig.1 (b) show the 1/M-vs-T plot of the data presented in fig.1 (a). The negative intercept of Curie-Weiss plot on temperature scale of the paramagnetic part has been clearly indicated. Fig.1(c,d) shows M-H behaviour of $La_{0.7}Bi_{0.3}CrO_3$ at (c) room temperature (RT) and (d) 10K. The perfect linearity of M-H curve shows that $La_{0.7}Bi_{0.3}CrO_3$ is paramagnetic at RT. The M-H taken at 10K shows a well defined hysteresis loop with non-saturating magnetization even up to field of 5T. The remnant magnetization and coercive field are determined to be 0.318 emu/gm and 2546 Oe respectively.

The occurrence of sharp rise of the magnetization in a FC M-T data, the non-saturating M-H loop, the small value of magnetization and the negative intercept of the Curie-Weiss plot indicate that $La_{0.7}Bi_{0.3}CrO_3$ under goes a paramagnetic to CAF type spin ordering at 230K. The undoped $LaCrO_3$ is know to be antiferomagnetic (AF)[2]. The occurrence of finite magnetization in $La_{0.7}Bi_{0.3}CrO_3$ could result from several possibilities. Keeping in view the high volatility of Bi at high temperatures, a few % of ‘A’ site vacancies may not be avoided. Such vacancies would lead to the presence of $Cr^{4+}$ together with $Cr^{3+}$, which may lead to double exchange interaction giving rise to net ferromagnetic moment. But orders of magnitude monotonous rise in the resistivity (data not shown) discards this possibility. The ferroelectric nature of $La_{0.7}Bi_{0.3}CrO_3$ [2], as shown below, even in a centrosymmetric structure (Pnma)[3], non-zero magnetic moment may appear due to the presence of anisotropy introduced by lone pair electron of doped Bi. The Reitveld refinement results clearly indicate anisotropic structural distortion as a function of Bi doping. Such distortions give possibility of non-zero ferromagnetic moment arising due to Dzyaloshinsky–Moriya (D–M) interaction as reported in the case of $BiFeO_3$[4]. The D–M interaction [4] arises due to asymmetric exchange interaction between the moments. The asymmetry in the interaction gives possibility other than that of the usual Heisenberg interaction, which gives ferromagnetic and antiferromagnetic ordering. The D-M interaction causes canting of the moments, giving rise to improper cancellation of moments in an otherwise antiferromagnetically ordered system. Thus D-M interaction induces weak ferromagnetism in a system.

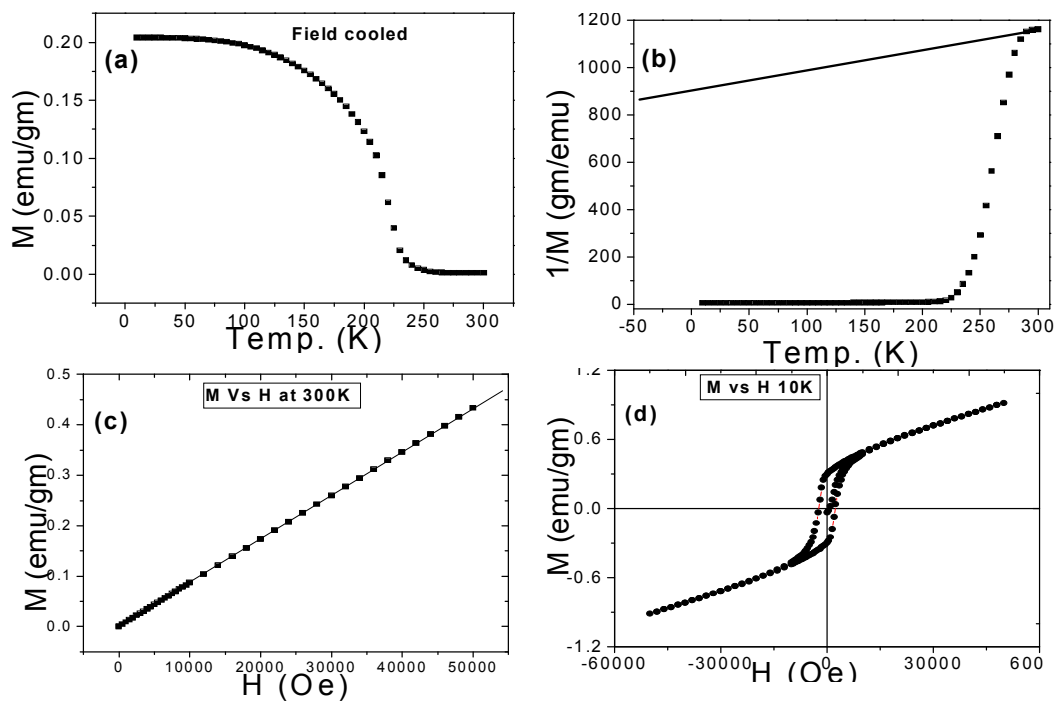

**Fig.1 (a).** 100 Oe FC Magnetization (M) vs Temperature (T) curve of $La_{0.7}Bi_{0.3}CrO_3$, (b) 1/M-vs-T plot of the same data. (c & d) M-H data taken at 300K and 10K respectively.

Fig.2(a) represents the ε′-T curves from 80K-390K at frequencies 1KHz, 10KHz , 100KHz and 1MHz. Fig.2(b) shows the enlarge portion of the same curve in fig. 2(a). From fig.2 (a) it can be seen that ε′-T variation shows pronounced frequency dispersion. At RT the ε′ decreases with increasing frequency. This is so because there is a strong frequency dispersion of ε′. Similar ε′-T behaviour has been reported for other similar multiferroic compounds like $YCrO_3$ [5] and $LuFe_2O_4$ [6].The frequency dispersion of ε′-T maximum towards high temperature is a signature of Debye type relaxation. Thus the present observation clearly shows that $La_{0.7}Bi_{0.3}CrO_3$ under goes a relaxor ferroelectric transition. Besides the high temperature relaxation, we observed another shoulder like variation at LT, with strong frequency dispersion. This shoulder appears at 103K, 119K, 143 and 180K for 1KHz, 10KHz, 100KHz and 1MHz frequencies respectively. We also observe another interesting feature i.e a sudden jump in the ε′-T data at 230K without any frequency dispersion. The jump in ε′-T coincides with $T_N$. The coincidence of jump in ε′-T variation and $T_N$ clearly shows the presence of ME-coupling in this material. The observed LT relaxation is identical to that of the $LuFe_2O_4$ [6], in which charge ordering (CO) of $Fe^{2+}$ and $Fe^{3+}$ ions takes place. Keeping in view the coexistence of $Cr^{+4}$ with $Cr^{+3}$ in the present ceramics too, the possibility of ordering of $Cr^{4+}$ and $Cr^{3+}$ may not be avoided. However in certain ceramics like $CaCu_3Ti_4O_{12}$[7], where very high dielectric constant, of the order of $10^5$ is observed due to " internal barrier layer capacitor effect" of the grain boundaries, such step like relaxation is attributed to Maxwell-Wagner(MW) type relaxation. But in the present case of $La_{0.7}Bi_{0.3}CrO_3$ the relative permittivity is not even 500, therefore MW relaxation may not be responsible. The presence of CO effect however remains a subject matter of further structural studies at LT.

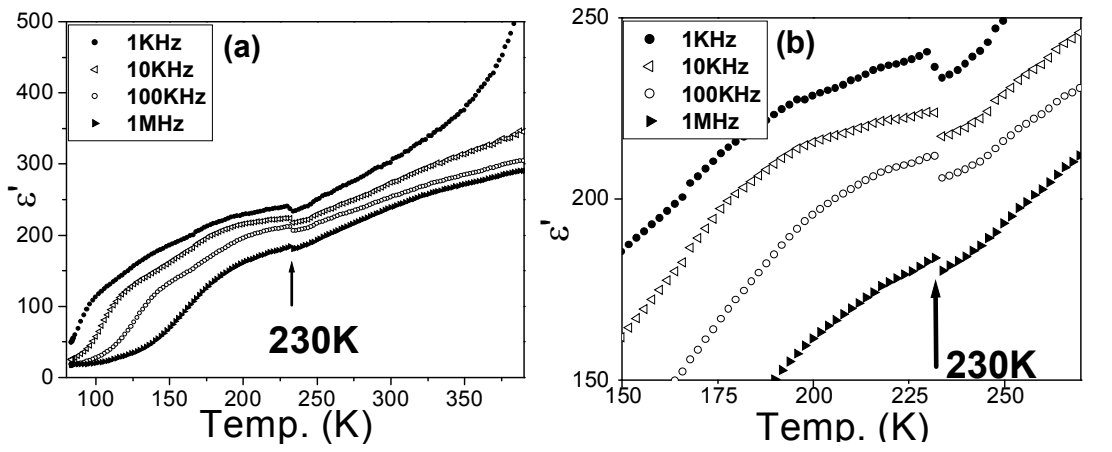

**Fig.2(a,b).** Temperature variations of the ε′ of $La_{0.7}Bi_{0.3}CrO_3$

## CONCLUSION

Based on the above studies it may be concluded that below 230K CAF and RFE orders both coexist in a single phase $La_{0.7}Bi_{0.3}CrO_3$ making it a multiferroic material. The observed anomaly in the ε′-T, coincident with the CAF transition at $T_N$, proves the presence of ME coupling in this material. The presence of D-M interactions appears to be responsible for the observed multiferroicity.

## ACKNOWLEGEMENT

Aga Shahee would like to acknowledge CSIR-India for financial support.